\PassOptionsToClass{10pt}{revtex4-2}
\documentclass[aps,prl,showpacs,floatfix,twocolumn,byrevtex,superscriptaddress]{revtex4-1}
\usepackage{amsmath}
\usepackage{amssymb}
\usepackage{amstext}
\usepackage{amsopn}
\usepackage{amsfonts}
\usepackage{amsxtra}
\usepackage[english]{babel}
\usepackage{graphicx}
\usepackage{bm}
\usepackage{multirow}
\usepackage{dcolumn}
\usepackage{color}
\usepackage{hyperref}
\usepackage{todonotes}
\usepackage{verbatim}
\usepackage{soul}
\usepackage{xcolor}
\usepackage{lineno}
\usepackage{siunitx}
\usepackage[version=3]{mhchem}
\date{\today}

\begin{document}

\title{Flat Band Induced Negative Magnetoresistance in Multi-Orbital Kagome Metal} 

\author{Jie Zhang}
\affiliation{Materials Science and Technology Division, Oak Ridge National Laboratory, Oak Ridge, Tennessee 37831, USA}
\author{T. Yilmaz}
\affiliation{Condensed Matter Physics and Materials Science Department, Brookhaven National Laboratory, Upton, New York 11973, USA}
\author{J. W. R. Meier}
\affiliation{Materials Science and Technology Division, Oak Ridge National Laboratory, Oak Ridge, Tennessee 37831, USA}
\author{J. Y. Pai}
\affiliation{Materials Science and Technology Division, Oak Ridge National Laboratory, Oak Ridge, Tennessee 37831, USA}
\author{J. Lapano}
\affiliation{Materials Science and Technology Division, Oak Ridge National Laboratory, Oak Ridge, Tennessee 37831, USA}
\author{H. X. Li}
\affiliation{Materials Science and Technology Division, Oak Ridge National Laboratory, Oak Ridge, Tennessee 37831, USA}
\author{K. Kaznatcheev}
\affiliation{Condensed Matter Physics and Materials Science Department, Brookhaven National Laboratory, Upton, New York 11973, USA}
\author{E. Vescovo}
\affiliation{Condensed Matter Physics and Materials Science Department, Brookhaven National Laboratory, Upton, New York 11973, USA}
\author{A. Huon}
\affiliation{Materials Science and Technology Division, Oak Ridge National Laboratory, Oak Ridge, Tennessee 37831, USA}
\author{M. Brahlek}
\affiliation{Materials Science and Technology Division, Oak Ridge National Laboratory, Oak Ridge, Tennessee 37831, USA}
\author{T. Z. Ward}
\affiliation{Materials Science and Technology Division, Oak Ridge National Laboratory, Oak Ridge, Tennessee 37831, USA}
\author{B. Lawrie}
\affiliation{Materials Science and Technology Division, Oak Ridge National Laboratory, Oak Ridge, Tennessee 37831, USA}
\author{R. G. Moore}
\affiliation{Materials Science and Technology Division, Oak Ridge National Laboratory, Oak Ridge, Tennessee 37831, USA}
\author{H. N. Lee}
\affiliation{Materials Science and Technology Division, Oak Ridge National Laboratory, Oak Ridge, Tennessee 37831, USA}
\author{Y. L. Wang} \email{yilinwang@ustc.edu.cn}
\affiliation{Hefei National Laboratory for Physical Science at Microscale, University of Science and Technology of China, Hefei, Anhui 230026, China}
\author{H. Miao} \email{miaoh@ornl.gov}
\affiliation{Materials Science and Technology Division, Oak Ridge National Laboratory, Oak Ridge, Tennessee 37831, USA}
\author{B. Sales} \email{salesbc@ornl.gov}
\affiliation{Materials Science and Technology Division, Oak Ridge National Laboratory, Oak Ridge, Tennessee 37831, USA}

\begin{abstract}
Electronic flat band systems are a fertile platform to host correlation-induced quantum phenomena such as unconventional superconductivity, magnetism and topological orders. While flat band has been established in geometrically frustrated structures, such as the kagome lattice, flat band-induced correlation effects especially in those multi-orbital bulk systems are rarely seen. Here we report negative magnetoresistance and signature of ferromagnetic fluctuations in a prototypical kagome metal CoSn, which features a flat band in proximity to the Fermi level. We find that the magnetoresistance is dictated by electronic correlations via Fermi level tuning. Combining with first principles and model calculations, we establish flat band-induced correlation effects in a multi-orbital electronic system, which opens new routes to realize unconventional superconducting and topological states in geometrically frustrated metals.

\end{abstract}

\maketitle

In correlated quantum materials, the importance of many-body effects to the ground state is primarily dictated by the relative energy scales between Coulomb interactions, $U$, and total electronic bandwidth, $W$. Recently, topological and symmetry-breaking orders are discovered in flat band Moire-superlattices, include Mott insulator \cite{Cao2018,Regan2020,Tang2020}, superconductivity \cite{Cao2018SC,Lu2019,Yankowitz1059}, ferromagnetism \cite{Yamamoto15704,Cao2020, Chen2020, Liu2020,Sharpe605, Shen2020} and quantum anomalous Hall effect \cite{Serlin900}. These observations triggered a resurgence of kinetic energy-driven correlation effects (see Fig. 1(a)) in both two-dimensional (2D) and bulk materials. The kagome lattice has been predicted to be an ideal flat band system \cite{Mielke1991,Tasaki1992, Sachdev1992}. In contrast to Moire-superlattices, which host isolated flat bands with \SI{}{\milli\eV} bandwidth\cite{Cao2018}, existing bulk kagome materials are multi-orbital systems, where the flat band coexists with eV-scale conduction electrons \cite{Liu2020CoSn,Meier2020,Kang2020FeSn,Kang2020CoSn}. The coupling among flat band electrons (heavy) with light electrons (those with large bandwidths), significantly elevates the quasi-particle coherence temperature and allows high-temperature electronic orders \cite{Kuroki2005,Matsumoto2018,Tang2011,Sheng2011,Neupert2011,Sun2011}. While various bulk kagome metals have been discovered \cite{Liu2020CoSn,Meier2020,Kang2020CoSn,Kang2020FeSn,Yin2020}, flat band-induced correlation effects, in particular transport and thermodynamic responses have yet to be discovered. In this letter, we combine transport, angle-resolved photoemission spectroscopy (ARPES) and theoretical calculations to demonstrate the flat band-controlled correlation effects in a prototypical kagome metal CoSn.

\begin{figure*}
\includegraphics[width=\textwidth]{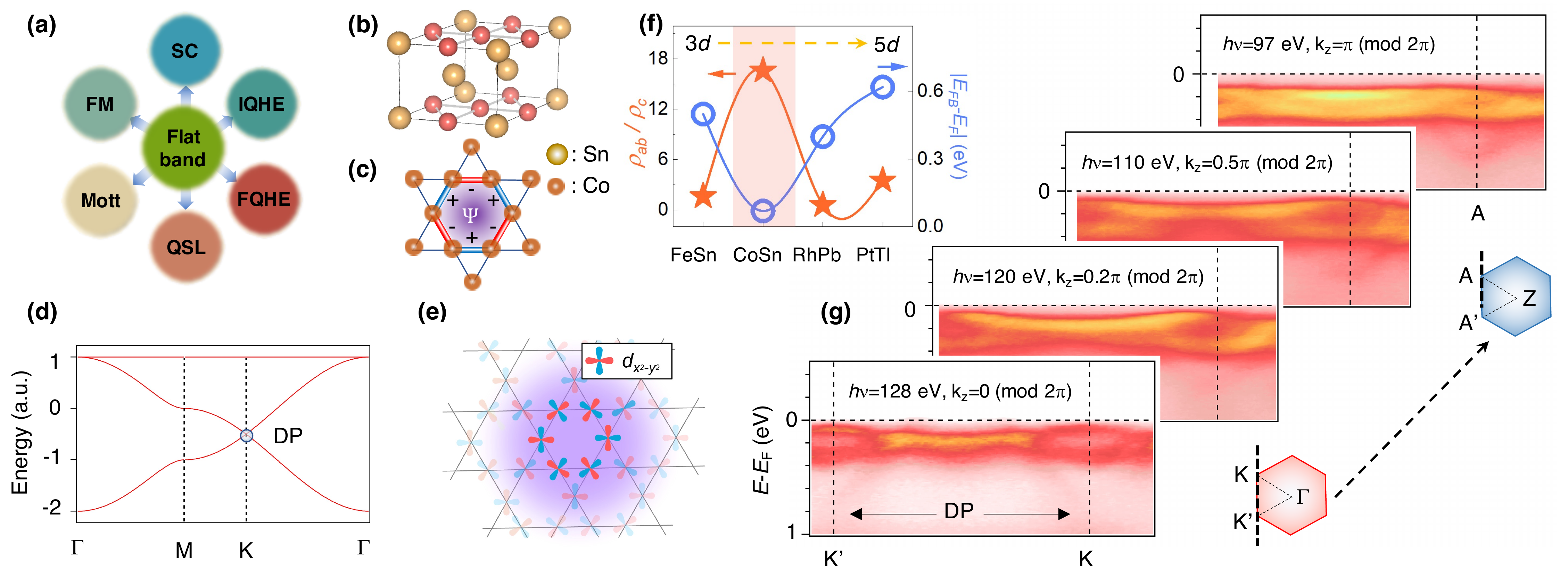}
\caption{Flat bands in kagome metal CoSn. (a) Diagram of possible flat band-induced topological and correlated quantum states—superconductivity (SC), integer/fractional quantum Hall effect (I/FQHE), quantum spin liquid (QSL), Mott insulator and ferromagnetism (FM). (b,c) show the crystal structure of CoSn and \ce{Co_{3}Sn} kagome layer, respectively. (d) The s-orbital and nearest neighbor tight binding model of the kagome lattice yields a flat band and a Dirac point (DP) at the K point. (e) Schematics showing the emergence of a flat band with the $d_{x^{2}-y^{2}}$-orbital in CoSn\cite{Kang2020CoSn}. (f) Extracted $|E_{FB}-E_{F}|$ (blue) and $\rho_{ab}/\rho_{c}$ (orange) of CoSn-type kagome metals \cite{Kang2020CoSn,Meier2020,Liu2020CoSn} indicate flat band-induced high resistivity anisotropy in CoSn. (g) Photon energy dependent ARPES intensity plots along the Brillouin zone boundary. Photon energies and their corresponding $k_{z}$ are shown in each panel.
}\label{fig1}
\end{figure*}

The CoSn-type structure, shown in Fig.~\ref{fig1}(b) and (c) can be built with 3$d$ (Fe, Co, Ni), 4d (Rh) or, 5d (Pt) transition metals on the kagome lattice \cite{Meier2020, Supp}. Theoretically, the flat band in kagome lattice arises from a simplified model that considers the nearest neighbor hopping between the isotropic s-orbitals, as illustrated in Fig.~\ref{fig1}(d). In CoSn, however, flat bands have non-zero band width due to imperfect interference of 3$d$ orbitals and hopping terms beyond the nearest neighbors, as illustrated in Fig.~\ref{fig1}(e). To confirm the presence of the flat band near the Fermi level $E_{F}$ in CoSn, we show the photon-energy dependent ARPES intensity plots along the high-symmetry $K-K^{’}$ and equivalent directions with different $k_{z}$ in Fig.~\ref{fig1}(g). At the K point, the flat band is \SI{70}{\milli\eV} below $E_{F}$, consistent with previous studies \cite{Kang2020CoSn,Liu2020CoSn}. In a single orbital Hubbard model, a flat band in proximity to $E_{F}$ is predicted to induce ferromagnetic fluctuations, which enhance electronic scattering in the kagome plane. Applying an external magnetic field suppresses low-energy magnetic fluctuations and gives rise to a negative magnetoresistance \cite{Yamada1972,Ueda1976,Arita2000}. Interestingly, as we show in Fig.~\ref{fig1}(f), the resistivity anisotropy, defined as $\rho_{ab}/\rho_{c}$ at \SI{2}{\kelvin}, is unusually large in CoSn. In comparison with other CoSn-type kagome metals \cite{Liu2020CoSn,Kang2020CoSn,Meier2020}, the energy difference between the top of the flat band and the Fermi level, $|E_{FB}-E_{F}|$, coincidentally reaches an maximum in CoSn, which suggests that the flat band enhances resistivity in the kagome plane.

\begin{figure*}
\includegraphics[width=0.9\textwidth]{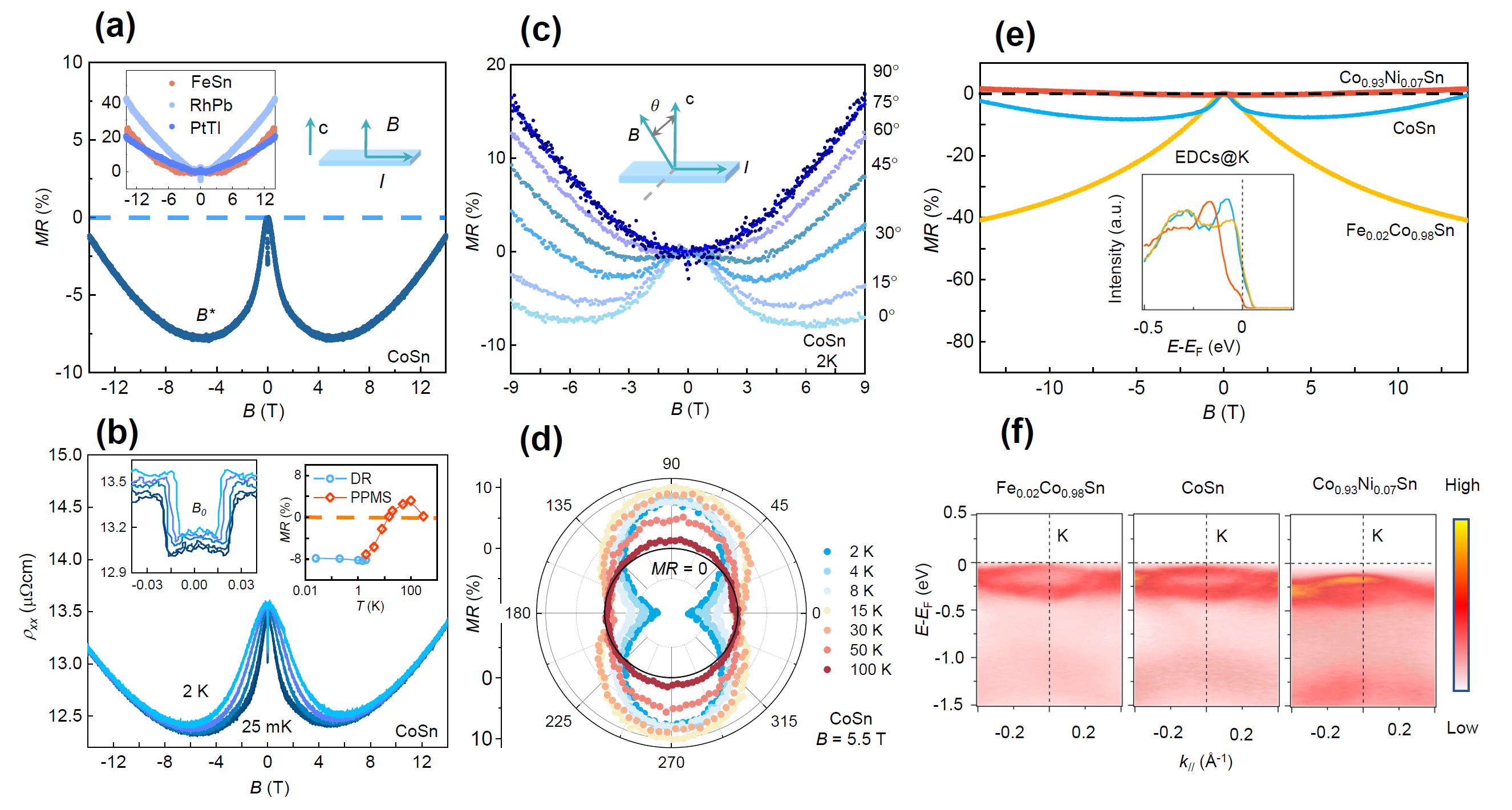}
\caption{Negative magnetoresistance and its doping-dependent evolution. (a) Magnetoresistance of CoSn-type kagome metals. Under the experimental geometry illustrated in the inset, only CoSn displays negative magnetoresistance. (b) Field-dependent resistivity across a range of temperatures. Left inset is a zoom-in of the same figure showing a low-field fine feature. Both the temperature and field dependence of the feature are consistent with the Sn flux-induced superconductivity. Right inset is the magnetoresistance at elevated temperatures demonstrating the transition from negative magnetoresistance below \SI{15}{\kelvin} to slightly positive magnetoresistance above \SI{30}{\kelvin}. (c) Large anisotropic magnetoresistance is observed when the magnetic field is rotated perpendicular to the current direction by $\theta$. $\theta$ is defined as the angle between the magnetic field and the crystal $c$-axis. (d) Angle-dependent magnetoresistance measured at $B^{*}$=\SI{5.5}{\tesla} shows that magnetoresistance anisotropy evolves as temperature increases. (e) Magnetoresistance of electron and hole doped CoSn. In electron-doped \ce{Co_{0.93}Ni_{0.07}Sn}, magnetoresistance is positive as expected for normal metals while in hole-doped \ce{Co_{0.98}Fe_{0.02}Sn}, the negative magnetoresistance is further enhanced up to $-40\%$ at \SI{14}{\tesla}. ARPES EDCs at the K point of \ce{(Fe{,}Co{,}Ni)Sn}. In \ce{Co_{0.98}Fe_{0.02}Sn}, the leading edge of EDC is shifted by about 40 meV at the K point. The ARPES intensity plots of \ce{(Fe{,}Co{,}Ni)Sn} near the K point are shown in (f).
}\label{fig2}
\end{figure*}

Figure~\ref{fig2}(a) shows magnetoresistances of all CoSn-type kagome metals at \SI{25}{\milli\kelvin}. While FeSn, RhPb and PtTl universally exhibit positive magnetoresistance as expected for a conventional metal, an unusual negative magnetoresistance is observed in CoSn. At high field, $B^{*}\sim$ \SI{5.5}{\tesla}, the magnetoresistance in CoSn shows a bend-back feature, in agreement with the coexistence of light and heavy electrons in CoSn\cite{Liu2020CoSn,Kang2020CoSn,Meier2020}. Fig.~\ref{fig2}(b) shows the magnetoresistance of CoSn down to \SI{25}{\milli\kelvin}. For $B < B^{*}$, the negative magnetoresistance is slightly enhanced at lower temperatures, while for $B>$ \SI{10}{\tesla}, all resistivity data collapse to a single curve. We note that the magnetoresistance shows a fine feature at very low field ($|B|< $\SI{0.03}{\tesla}), which we attribute to the superconducting transition of Sn-flux. Above \SI{2}{\kelvin}, the magnetoresistance shows two temperature scales: a negative magnetoresistance temperature below \SI{15}{\kelvin} and a positive magnetoresistance temperature up to \SI{100}{\kelvin}, as summarized in Fig. 2(b) inset. These two temperature scales feature distinct magnetoresistance-anisotropy, MR($\theta$), where $\theta$ is defined as the angle between the magnetic field and the $c$-axis of the kagome structure.
As shown in Fig.~\ref{fig2}(c) and (d), at \SI{2}{\kelvin} MR($\theta$) displays a sign changes from negative to positive as $\theta$ moves from 0 to 90 degree.
Increasing temperature monotonically reduces the anisotropy and it becomes nearly isotropic above \SI{100}{\kelvin}. The sign-change MR($\theta$) at lower temperature suggests large electronic anisotropy. This is consistent with the flat band enhanced ferromagnetic fluctuations in the kagome plane, where at $\theta$=0 and 90 degree the cyclotron orbits are in- and out- of the kagome plane.

With the negative magnetoresistance and large $\rho_{ab}/\rho_{c}$ established in CoSn, we further prove that these transport phenomena can be effectively controlled by tuning $|E_{FB}-E_{F}|$ through chemical doping. Figure~\ref{fig2}(e) compares the magnetoresistance of the hole-doped \ce{Co_{0.98}Fe_{0.02}Sn} and electron-doped \ce{Co_{0.93}Ni_{0.07}Sn}. The chemical potential shift is corroborated by the ARPES intensity plots in \ce{Co_{0.98}Fe_{0.02}Sn}, CoSn and \ce{Co_{0.93}Ni_{0.07}Sn}, where the leading edge of the energy distribution curves (EDCs) at the K point (inset of Fig.~\ref{fig2}(e)) moves by +\SI{40}{\milli\eV} in \ce{Co_{0.98}Fe_{0.02}Sn} and -\SI{100}{\milli\eV} in \ce{Co_{0.93}Ni_{0.07}Sn}. Remarkably, the negative magnetoresistance is significantly enhanced in \ce{Co_{0.98}Fe_{0.02}Sn} but changes sign in \ce{Co_{0.93}Ni_{0.07}Sn}. We shall note, however, that the Fe doping may have a non-trivial impurity effect. As we show in the inset of Fig.~\ref{fig2}(e), comparing with CoSn, the \SI{70}{\milli\eV} quasi-particle peak in \ce{Co_{0.98}Fe_{0.02}Sn} is significantly suppressed while the \SI{300}{\milli\eV} peak remains nearly unchanged. Assuming a rigid-band shift at low-doping level, this observation supports enhanced low-energy electronic scattering, which consequently, enhances the density of state \cite{Sales2021} at $E_{F}$ due to a broadened low-energy spectral function. As we will demonstrate below, CoSn is near the boundary of magnetic ordering with ferromagnetic fluctuations in the kagome plane. Fe-doping may thus polarize its surrounding electronic state and form isolated magnetic clusters with spin-glass behavior at low temperature \cite{Sales2021}. Such an impurity effect is expected to be weak or absent in the electron doped \ce{Co_{0.93}Ni_{0.07}Sn}, where the magnetic fluctuations are suppressed.

\begin{figure*}
\includegraphics[width=0.9\textwidth]{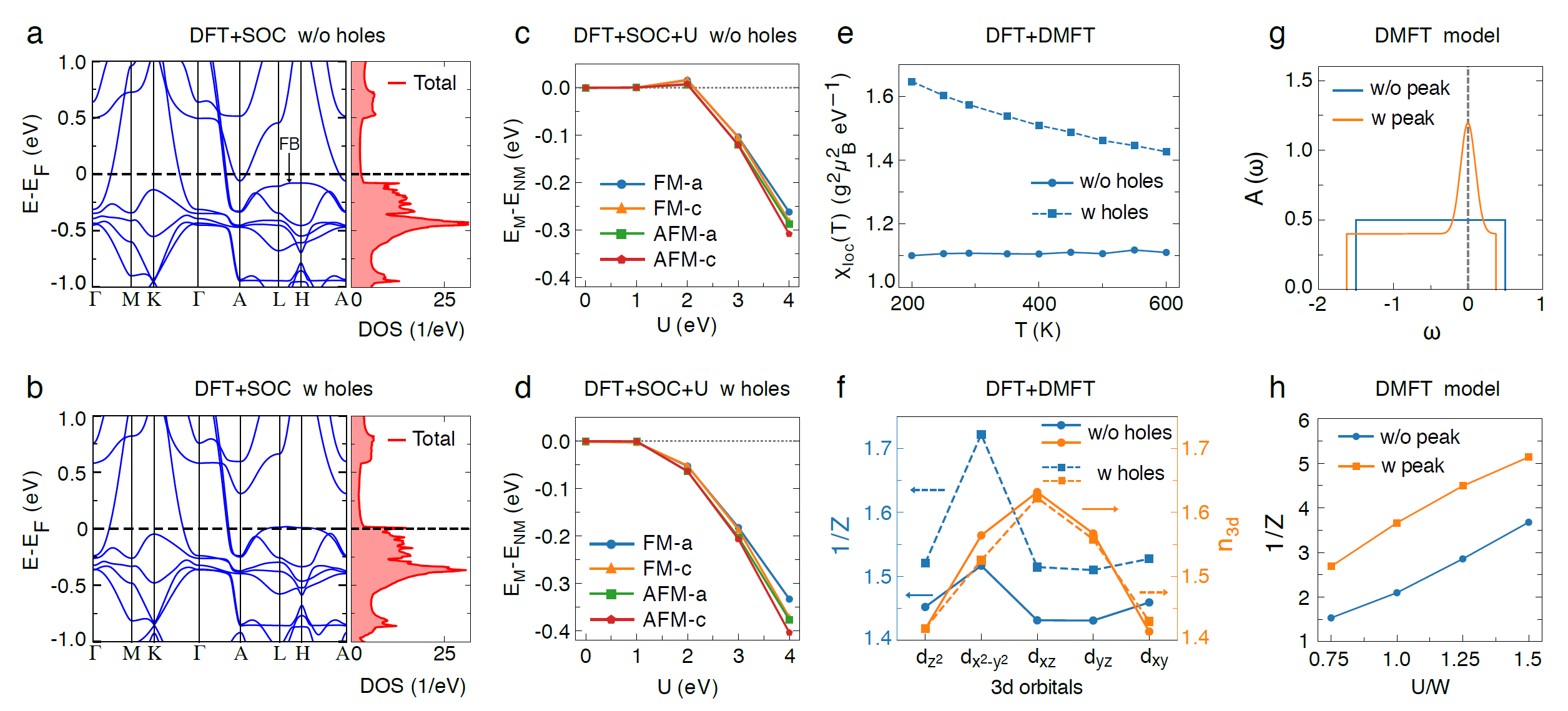}
\caption{Flat band-enhanced electronic correlations and magnetic fluctuations in CoSn. (a,b) The DFT+SOC calculated band structures and total density of states for CoSn without (a) and with (b) holes, respectively. In (b), 0.5 holes per unit cell are added to push the Fermi level close to the flat band. (c,d) The energy of magnetic states relative to non-magnetic states calculated by DFT+SOC+U as functions of Hubbard $U$, for CoSn without (c) and with (d) holes, respectively. (e) Local static magnetic susceptibilities as function of temperature $T$. (f) Orbital-resolved quasi-particle mass enhancement and occupancy numbers calculated by DFT+DMFT at $U=$ \SI{5}{\eV}, $J_{H}=$ \SI{0.8}{\eV} and  $T=$\SI{290}{\kelvin}. (g) Non-interacting density of states $A(\omega)$ used in the DMFT calculations of a degenerate five-orbital Hubbard model. One is flat (blue) while another one is augmented with a Gaussian type peak at the Fermi level to simulate the effect of a flat band (orange). The Fermi level is chosen to make the total occupancy 7.5, a value that is close to the 3d occupancy of CoSn. (h) Quasi-particle mass enhancement of the five-orbital Hubbard model as a function of $U/W$. Results are obtained from DMFT calculations at Hund’s coupling $J_{H}=$\SI{0.16}{U} and temperature $k_{B}T=$\SI{0.01}{W}, where $W$ is the non-interacting band width.
}\label{fig3}
\end{figure*}

To understand the experimental observations, we perform density functional theory plus spin-orbital coupling (DFT+SOC) and DFT plus dynamical mean-field theory (DFT+DMFT) calculations. Figure~\ref{fig3}(a) and (b) show the DFT+SOC band structure without and with 0.5-hole (per unit cell) doping in CoSn. In the latter case, the flat band is pushed to $E_{F}$. Figure~\ref{fig3}(c) and (d) present the energy difference between magnetic and non-magnetic ground states calculated by DFT+SOC+U as functions of Hubbard U for without and with hole-doped CoSn, respectively. We compare the energy of four magnetic configurations to the non-magnetic state. We considered ferromagnetism (FM) and A-type antiferromagnetism (AFM) with moments parallel to the hexagonal $c$ and $a$ axis (-$c$ and -$a$), respectively. For large $U$, the AFM-$c$ is the ground state for both without and with hole doped CoSn, however the critical $U$ for onset of magnetic ordering is significantly smaller for with hole-doped CoSn. This facilitates stronger magnetic fluctuations in the kagome plane. This observation is further confirmed by the DFT+DMFT calculations shown in Fig. 3(e) and (f), where the local static magnetic susceptibility, $\chi_{loc}(T)$, displays a Pauli to Curie-Weiss crossover. Meanwhile, the quasi-particle mass enhancement, $1/Z$, is enhanced for all 3$d$-orbitals, from without to with hole-doped CoSn. As an important aspect, we show by the orange curves in Fig. 3(f) that the doped holes primarily reside on the Sn $p$-orbitals, which leaves the occupation number of 3$d$-orbitals nearly fixed (about a reduction of 0.05 electrons per Co). This finding strongly suggests that the enhanced correlation effects are mainly due to the proximity of the flat band to $E_{F}$, instead of 3$d$ occupancy reduction (see Supplementary Materials).

\begin{figure}
\includegraphics[width=8 cm]{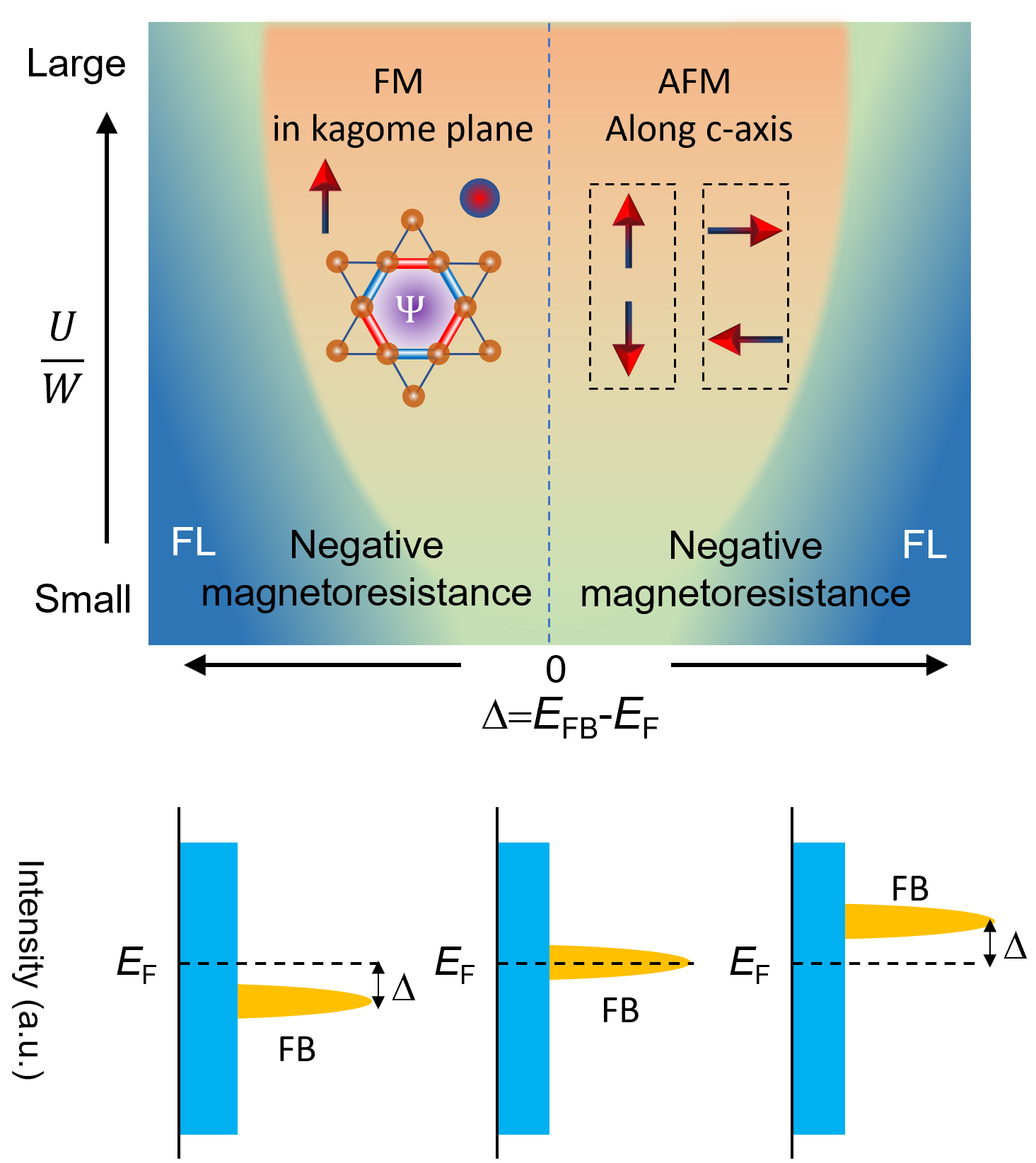}
\caption{Schematic phase diagram of bulk flat band materials. The vertical and horizontal axis are $U/W$ and $\Delta=E_{FB}-E_{F}$, respectively. With large $U/W$ and $\Delta \ll W$, the electronic state tends to be magnetically ordered. However, unlike 2D systems, where ferromagnetic (FM) order is the groun state, in bulk flat band systems, the ground state energy is highly dependent on the interlayer magnetic interactions. In CoSn-type kagome metals, the A-type AFM order is energetically favored, although the spins are still ferromagnetically aligned within the kagome layers. When $U/W\sim1$ and $\Delta \to 0$, such as CoSn and \ce{Co_{0.98}Fe_{0.02}Sn}, the system is near the boundary of magnetically ordered phase. The flat band will induce large ferromagnetic fluctuations in the kagome layer and yield anisotropic negative magnetoresistance (indicated by yellow regions). In the case of $U/W \ll 1$ and $|\Delta| \sim W$, the system is expected to behave as a good metal with Fermi liquid (FL) behavior.
}\label{fig4}
\end{figure}

The generality of flat band enhanced correlation effects in multi-orbital systems is confirmed by a simple DMFT model study, where we consider a degenerate five-orbital Hubbard model with (orange) and without (blue) the flat band near $E_{F}$ (Fig.~\ref{fig3}(g)). As we show in Fig.~\ref{fig3}(h) and the supplementary materials, while $1/Z$ is generally enhanced by increasing $U/W$, placing the flat band near $E_{F}$, progressively enlarges $1/Z$ even with a relatively small $U/W$. 

Our experimental and theoretical results lead to a schematic phase diagram for bulk multi-orbital flat band systems, as shown in Fig.~\ref{fig4}. The vertical and horizontal axis are $U/W$ and $\Delta=|E_{FB}-E_{F}|$, respectively. The key information is that the presence of flat band near $E_{F}$ progressively enhances correlation and ferromagnetic fluctuations in the kagome plane even in relatively small $U/W$ regime. Unlike 2D single-orbital flat band systems, where ferromagnetism is the ground state \cite{Mielke1991,Tasaki1992}, the coexistence of light electrons and flat band plays an important role for interlayer couplings that lead to magnetic ordering. Near the boundary of a magnetic long-range order, which is the case of CoSn, the flat band will induce ferromagnetic fluctuations in the kagome plane and give rise to anisotropic negative magnetoresistance. In this critical region, magnetic fluctuations can be controlled by tuning $\Delta$ providing an efficient way to control correlation effects in multi-band systems.

Acknowledgements: 
H. M. thank G. Kotliar and B. H. Yan for stimulating discussions. This research was sponsored by the U.S. Department of Energy, Office of Science, Basic Energy Sciences, Materials Sciences and Engineering Division (Transport, Synthesis and Bulk Characterizations) and by the Laboratory Directed Research and Development Program of Oak Ridge National Laboratory, managed by UT-Battelle, LLC, for the U. S. Department of Energy (ARPES). The DFT and DFT+DMFT calculations were performed on ThianHe-1A, the National Supercomputer Center in Tianjin, China. J. Zhang and T. Yilmaz contributed equally to this work.

\bibliography{ref}

\end{document}